\documentclass[12pt,preprint]{aastex}
\usepackage{graphicx}

\begin{document}

\title{The Distance, Mass, and Radius of the Neutron Star in 4U\,1608$-$52}

\author{Tolga G\"uver$^{1}$, Feryal \"Ozel$^{1}$, Antonio
  Cabrera-Lavers$^{2,3}$ and Patricia Wroblewski$^{1}$}

\affil{$^{1}$ Department of Astronomy, University of Arizona, 933
  North Cherry Avenue, Tucson, AZ,USA}

\affil{$^{2}$ Instituto de Astrofisica de Canarias, E-38205 La Laguna,
  Tenerife, Spain}

\affil{$^{3}$ GTC Project Office, E-38205 La Laguna, Tenerife, Spain}

\begin{abstract}

  Low mass  X-ray binaries (LMXBs) that show  thermonuclear bursts are
  ideal sources for constraining the equation of state of neutron star
  matter. The  lack of independent  distance measurements for  most of
  these  sources, however,  prevents a  systematic exploration  of the
  masses  and   radii  of  the  neutron  stars,   hence  limiting  the
  equation-of-state  studies.  We  present here  a measurement  of the
  distance to the LMXB 4U\,1608$-$52 that is based on the study of the
  interstellar  extinction  towards the  source.  We  first model  the
  individual  absorption  edges of  the  elements  Ne  and Mg  in  the
  high-resolution  X-ray spectrum obtained  with XMM-Newton.   We then
  combine this information with a  measurement of the run of reddening
  with distance using red clump stars and determine a minimum distance
  to  the   source  of  3.9~kpc,   with  a  most  probable   value  of
  5.8~kpc. Finally,  we analyze time-resolved X-ray  spectra of Type-I
  X-ray bursts observed  from this source to measure  the mass and the
  radius  of  the neutron  star.   We  find a  mass  of  M$= 1.74  \pm
  0.14$~M$_{\sun}$ and a radius  of R$= 9.3 \pm 1.0$~km, respectively.
  This  mass  and radius  can  be  achieved  by several  multi-nucleon
  equations of state.
\end{abstract}

\keywords{stars: neutron - X-ray: individual (4U\,1608$-$52)}

\section{Introduction}

The potential of utilizing low-mass X-ray binaries (LMXBs) that show
thermonuclear X-ray bursts to measure the masses and radii of their
neutron stars has long been recognized (see, e.g., van Paradijs 1978,
1979; Lewin, van Paradijs, \& Taam 1993). During these short-lived
flashes, bright thermal emission from the surface of the neutron star
is observed, allowing, in principle, a measurement of the stellar
radius or its surface gravity. Indeed, a number of methods that
involve the luminosity, surface redshift, or flux oscillations during
these bursts have been proposed and used to measure or constrain the
masses and radii of neutron stars in LMXBs (for a recent review, see
Lattimer \& Prakash 2007). Here, we focus on the time-resolved X-ray
spectroscopy of thermonuclear X-ray bursts observed from 4U~1608$-$52
in order to measure both the mass and the radius of its neutron star.

In a subset of thermonuclear bursts called photospheric radius
expansion (PRE) bursts, the source flux reaches a peak that is thought
to correspond to the local Eddington flux on the stellar surface,
which is related to the mass and the radius of the neutron star. A
second spectroscopic quantity that can be used to infer the mass and
the radius of the neutron star is observed during the cooling tails of
thermonuclear bursts. During this cooling phase, the measured apparent
area of the emitting region remains constant while both the flux and
the color temperature decrease. It has been shown that the observed
surface areas remain constant between different bursts of the same
source (Galloway et al. 2008a), indicating that the emission arises
from a reproducible area on the stellar surface, likely corresponding
to the entire neutron star surface. In addition to these two
spectroscopic quantities, determining both the neutron star mass and
its radius requires an independent distance or gravitational redshift
measurement (see, e.g., \"Ozel 2006; \"Ozel, G\"uver, \& Psaltis 2009;
for a review of earlier results, see Lewin et al.\ 1993).

Apart from sources in globular clusters, independent measurements of
distances to LMXBs have been a long-standing challenge (see, e.g., van
Paradijs \& White 1995). With the advances in high-resolution X-ray
grating spectroscopy, it is now possible to determine the X-ray
absorption towards LMXBs with enough precision to allow studies of the
properties of interstellar matter (ISM) towards these sources (see,
e.g., Juett, Schulz, \& Chakrabarty 2004; Juett et al. 2006;
Wroblewski, G\"uver, \& \"Ozel 2008). Extinction measurement can in
principle allow for a distance measurement if combined with a good
standard candle within the field of view to an X-ray binary. Red clump
stars have shown to be such standard candles (Paczynski \& Stanek
1998; L\'opez-Corredoira et al. 2002). They are core helium-burning
giants that form a well-defined concentration of stars in a
color-magnitude diagram (CMD). Because their luminosities are largely
independent of the total stellar mass and their infrared colors are
insensitive to metallicity, they can be used to measure the run of
reddening with distance. Finding the distances to LMXBs, then,
involves a comparison of the extinction measurement of the X-ray
source to that of red clump stars in the same field of view. A very
similar technique has already been successfully applied to anomalous
X-ray pulsars by Durant \& van Kerkwijk (2006a, 2006b).

In this study, we measure the distance to the LMXB 4U\,1608$-$52 using
the red clump method in order to ultimately determine the mass and the
radius of its neutron star. In detail, we {\it(i)} use high resolution
X-ray spectroscopy to measure the column density in metals by modeling
the absorption  edges of  Mg and Ne;  {\it (ii)} use  the interstellar
abundance model  of Wilms,  Allen, \& McCray  (2000) to  determine the
amount of equivalent hydrogen  column density towards the source; {\it
  (iii)} use observed  relations between $N_{\rm H}-A_V$ (G\"orenstein
1975;  G\"uver \& \"Ozel  2009) and  $A_V-A_K$ (Cardelli,  Clayton, \&
Mathis  1989) to  convert  to infrared  extinction  $A_K$; {\it  (iv)}
create  a near-IR  CMD of the  field  towards the
X-ray binary to determine the run of reddening with distance using red
clump stars;  and {\it (v)}  compare the infrared extinction  $A_K$ to
the source  with this  run of reddening  with distance to  measure the
distance to the  X-ray binary. Finally, we {\it  (vi)} fit X-ray burst
spectra  to   measure  the  Eddington   flux  and  apparent   area  of
4U\,1608$-$52  and  combine these  with  the  distance measurement  to
derive the mass and radius of the neutron star.

4U\,1608$-$52 is a transient X-ray burster that was first detected by
the two Vela-5 satellites (Belian, Conner, \& Evans 1976). Subsequent
Uhuru observations confirmed the connection between the bursting and
the persistent source (Tananbaum et al. 1976). A more precise X-ray
position, obtained with HEAO-1, permitted the identification of the
optical counterpart, QX Normae, with an I=18.2 mag (Grindlay \&
Liller 1978). Wachter et al. (2002) identified a modulation in the
optical light curve of the source with a period of 0.537 days, which
they attributed to the orbital period. 4U\,1608$-$52 has also been
observed during several outbursts with the Rossi X-ray Timing Explorer
(RXTE). Thirty-one type-I X-ray bursts were identified, out of which 12 were
categorized as PRE bursts. During several PRE bursts, burst
oscillations have been detected with a frequency of 619 Hz, making
this source one of the most rapidly rotating accreting neutron-stars
(Galloway et al. 2008a).

In \S 2, we measure the galactic hydrogen column density towards the
source using absorption edges of individual elements. We measure the
distance to the source in \S 3, using its hydrogen column density in
conjunction with the red clump stars. In \S 4, we present results of
the time-resolved spectral modeling of Type-I X-ray bursts. In \S 5,
we use the distance and the results from the burst modeling to
determine the mass and the radius of the neutron star. Finally, we
discuss our results and its implications in \S 6.

\section{The Hydrogen Column Density Towards 4U\,1608$-$52}

In this study, we make use of the hydrogen column density towards
4U\,1608$-$52 in two ways: $(i)$ to measure the distance to the X-ray
binary and $(ii)$ to model the time-resolved X-ray spectra obtained
with the RXTE.  We, therefore, begin our analysis with a measurement
of the galactic X-ray absorption from high-resolution X-ray spectra.

We determine the amount of the interstellar X-ray absorption by
modeling the individual absorption edges of the elements Ne and Mg in
the X-ray grating spectra of the source obtained with the Reflection
Grating Spectrometer (RGS) onboard XMM-Newton.  This approach does not
rely on an assumption of the intrinsic broadband X-ray spectrum of the
source and is similar to the method previously applied to a number of
LMXBs by Juett et al. (2004, 2006) and Wroblewski et
al. (2008) and to anomalous X-ray pulsars by Durant \& van Kerkwijk
(2006a).

4U\,1608$-$52 has been observed 4 times with XMM-Newton. A list of
these observations together with the exposure time is given in
Table~\ref{xmm_obs}. The source was in a quiescent state during three
of the four observations, rendering very low counts. In addition, due
to the high energy particle background during the last observation
(Obs ID : 0149180201), only 5~ks of this observation were usable. Even
though 4U~1608$-$52 was in an outburst during this observation, the
short exposure time as well as the intrinsic spectrum of the binary
prevented us from measuring the optical depths of the Si, O, and Fe
edges. Future grating observations of this source during an outburst
phase would also enable a measurement of these edges.

We extracted RGS spectra with the latest calibration files available
(as of 2009 March 23) and used the {\it rgsproc} tool within SAS
v.7.1.0. By default, the binning of X-ray grating spectra oversamples
the instrumental resolution of the RGS detectors. For this reason, we
rebinned the resulting spectra by a factor of 5 to match the
wavelength resolution of the detector (0.04 \AA). The resulting X-ray
spectrum had approximately 300 counts at the Mg edge and 40 counts at
the Ne edge per spectral bin. To analyze the spectra, we used XSPEC
v12 (Arnaud 1996).
 
We modeled the Ne and Mg edges by fitting the X-ray spectrum in small
wavelength intervals: the 8.5$-$10.5~\AA region for Mg and the 
13.0$-$16.0~\AA region for Ne. We assumed that the spectrum in each of these
intervals can be modeled with a power-law function ($F_\lambda \propto
[\lambda/\lambda_{\rm edge}]^{\alpha}$) and modeled each edge with a
function of the form

\begin{eqnarray}
F_{\lambda} = \left \{
\begin{array}{lr}
F_{\lambda} & {\rm for~} \lambda > \lambda_{\rm edge} \\ F_{\lambda}
\exp\left[-\rm{A} \left( \frac{\lambda}{\lambda_{\rm edge}} \right)^3
\right] & {\rm for~} \lambda \leq \lambda_{\rm edge}.
\end{array} \right.
\end{eqnarray}

\begin{figure*}
\centering
\includegraphics[scale=0.40]{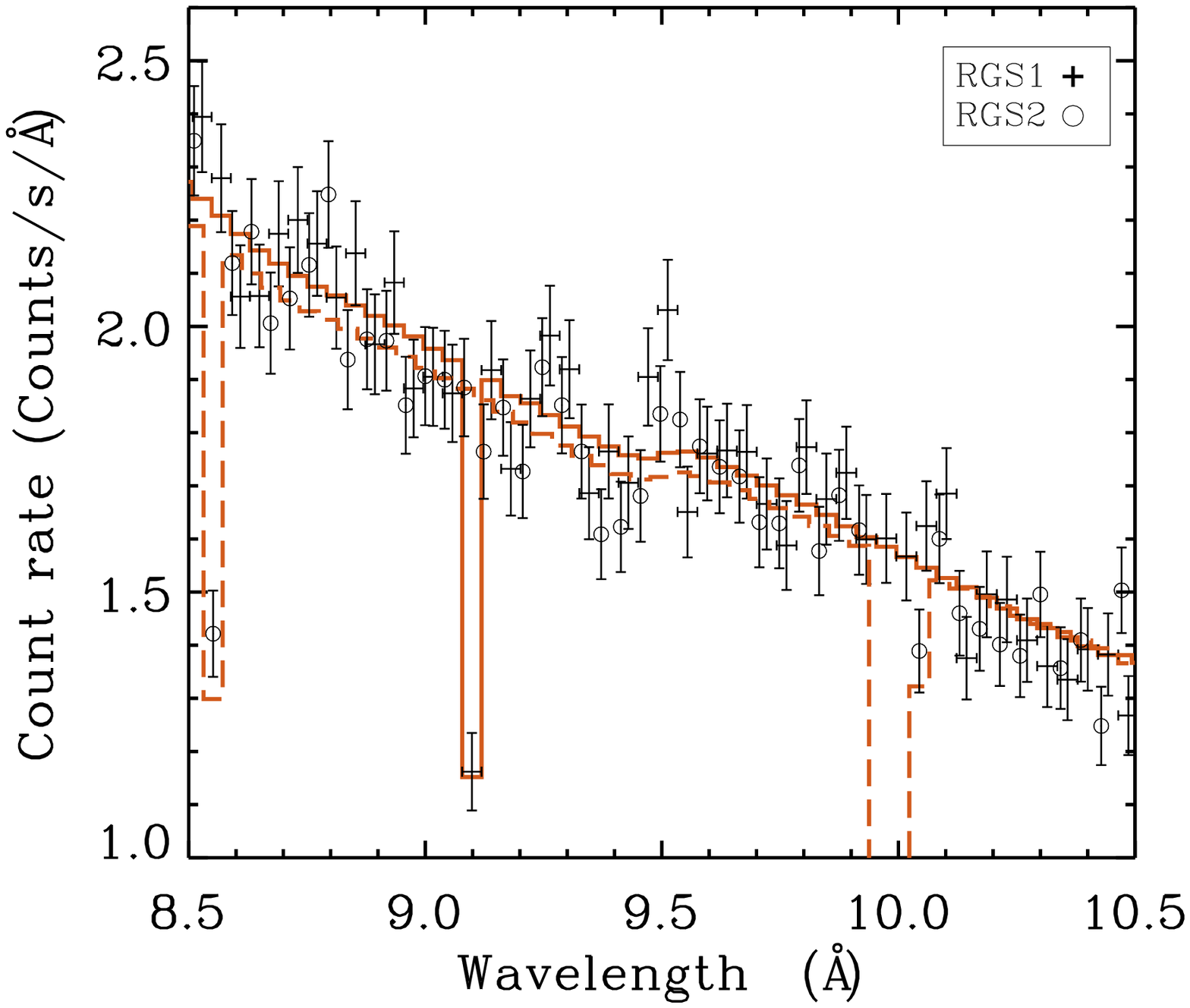}
   \includegraphics[scale=0.40]{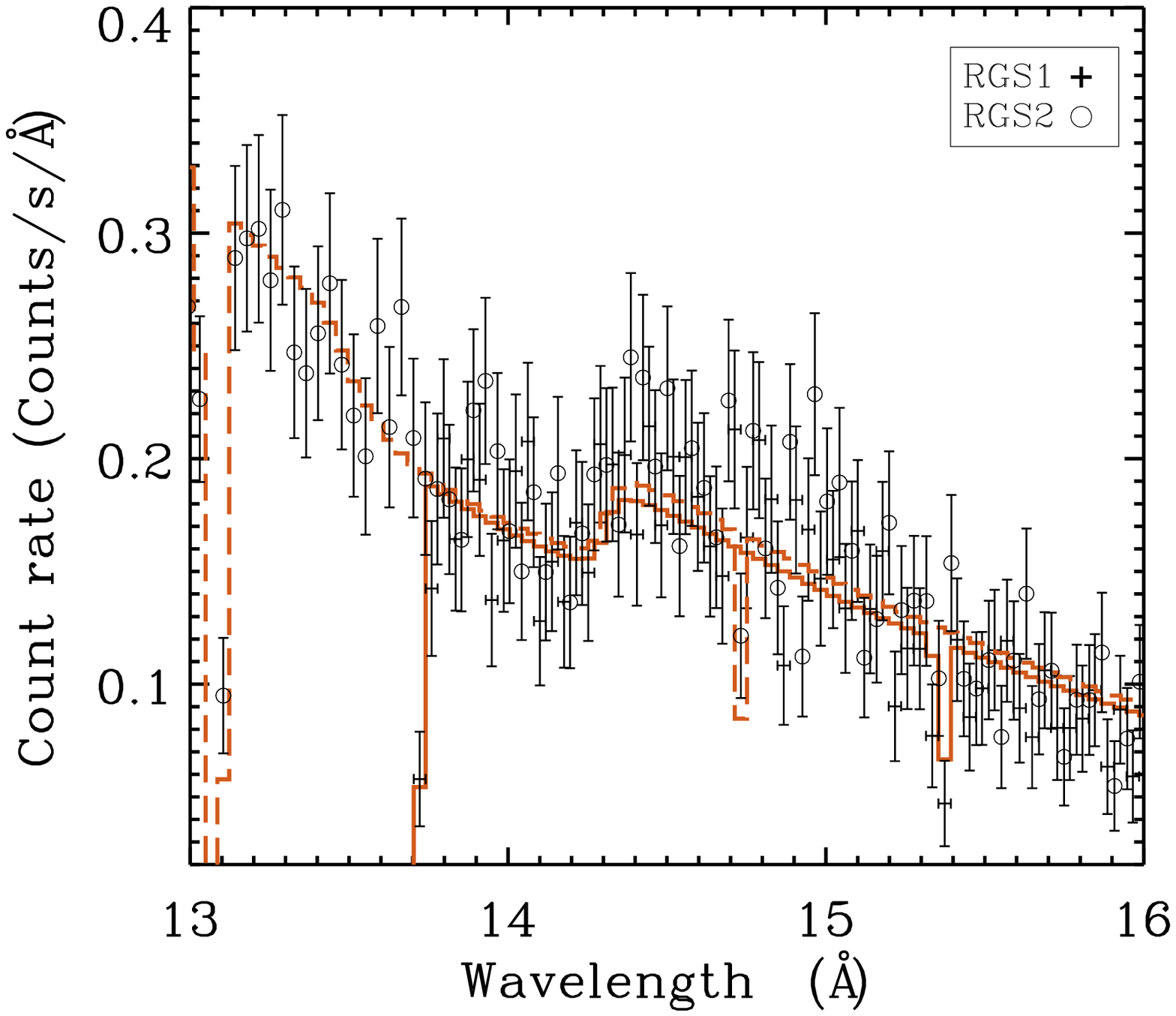} 
   \caption{RGS1 and RGS2 spectra of 4U\,1608$-$52 together with the
   best fit models (solid line for RGS1 and dashed line for RGS2) are
   given for the regions that are used to determine the column
   densities of Mg (left panel) and Ne (right panel). Sharp features
   are due to CCD gaps and the malfunctioning CCD in the RGS
   detector.}
\label{edges}
\end{figure*}

There are uncertainties in the positions of edges that are comparable
to the resolution of the detector (Juett et al.\ 2004).  Because of
the relatively low signal-to-noise of the data, leaving the edge
positions as free parameters did not affect the results of the
fits. We, therefore, fixed the edge positions at the values inferred
from high-resolution X-ray spectroscopy of X-ray binaries; i.e.,
$\lambda_{\rm edge}$ = 9.5~$\rm\AA$ for Mg (Ueda et al.\ 2005) and
14.3~$\rm\AA$ for Ne (Juett et al.\ 2004). We show the spectrum and
the best-fit models in Figure~\ref{edges} and summarize the results of
the fits in Table~\ref{edgeres}. Using the best-fit absorption
coefficients and the cross-sections from Gould \& Jung (1991), we then
determined the column densities of each element. Our best-fit values
are (9.65 $\pm$1.56) $\times10^{17}$~cm$^{-2}$ for $N_{\rm Ne}$ and
(2.42 $\pm$0.87) $\times10^{17}$~cm$^{-2}$ for $N_{\rm Mg}$. Here and
throughout this paper, the errors denote 1$- \sigma$ uncertainties.

Using these values and assuming the ISM abundances reported by Wilms
et al. (2000), we inferred a Galactic hydrogen column density towards
4U\,1608$-$52 of ($1.08 \pm 0.16) \times 10^{22}$~cm$^{-2}$.  We note
here that the hydrogen column density inferred from an analysis of the
continuum spectrum of 4U\,1608$-$52 obtained from EXOSAT observations
is (1.0$-$1.5) $\times 10^{22}$~cm$^{-2}$ (Penninx et al. 1989).

\subsection{Conversion from Hydrogen Column Density to Infrared 
Extinction}

Comparing the interstellar extinction determined from the X-ray
grating spectrum with the K band extinction of red clump stars (see \S
3) requires converting the hydrogen column density to an infrared
extinction $A_K$. We carry out this conversion in two steps using
empirical laws that stem from observations of large samples of
sources. We first convert from $N_{\rm H}$ to optical extinction $A_V$
and subsequently from $A_V$ to $A_K$.

The observed relation between hydrogen column density and optical
extinction has been the subject of numerous studies (e.g.,
G\"orenstein 1975; Predehl \& Schmitt 1996; G\"uver \& \"Ozel 2009).
Using a variety of X-ray sources such as supernova remnants or X-ray
binaries, the relation between $N_{\rm H}$ and $A_V$ has been measured
along different lines of sight. All of the studies found a relatively
tight linear relationship between these quantities and none revealed
any significant variation along different lines of sight.

Here, we use the relation  

\begin{equation}
N_{\rm{H}} = (2.21 \pm 0.09) \times10^{21}\times A_{\rm{V}}. 
\end{equation}

(see G\"uver \& \"Ozel 2009 and references therein) to convert the
hydrogen column density to optical extinction. This result comes from
the most recent effort to quantify the A$_{V}-$N$_{H}$ relation and is
based on high-resolution spectra of 22 supernova remnants. The
uncertainty in the normalization incorporates the uncertainties in the
measurements of both quantities for each source and reflects also the
variation along different lines of sight that is consistent with the
data. This is why the uncertainty in the normalization is larger
compared to previous studies despite the higher quality data used and
the larger number of sources. Using this conversion law, we found an
optical extinction towards the X-ray binary of $4.89 \pm 0.75$~mag.

We carried out the second conversion between optical and infrared
extinction following the empirical relation of Cardelli, Clayton, 
\& Mathis (1989). Using a large sample of stars, these authors 
showed that {\it (i)} the extinction law is highly uniform for
wavelengths greater than 0.7 microns (where the K band lies), {\it
(ii)} the extinction law depends very weakly on $R_V$ in this
wavelength range, and {\it (iii)} it has extremely little scatter
along different lines of sight. The ratio $A_K/A_V$ is found
empirically to depend on the parameter $R_V$ according to the relation
\begin{equation}
\frac{A_K}{A_V} = 0.1615 - \frac{0.1483}{R_V}. 
\end{equation}
Allowing the parameter $R_V^{-1}$ to take values in the range
$0.2-0.4$, as in the observed sample of Cardelli et al. (1989), we
find that the ratio $A_K/A_V$ ranges between $0.10-0.13$. We assume a
box-car distribution for the ratio $A_K/A_V$ within this range. We
found an extinction of $A_K = 0.56 \pm 0.11$~mag towards
4U\,1608$-$52.

\section{Determination of the Distance to 4U\,1608$-$52 by Means of
  Red Clump Stars}

In order to determine the distance to 4U\,1608$-$52, we use a
technique that relies on the red clump stars that are present in the
field of view of the source. This method, discussed in detail in
L\'opez-Corredoira et al. (2002) and Cabrera-Lavers, Garz\'on, \&
Hammersley (2005) is very reliable in deriving both the stellar number
density and the interstellar extinction along a given line of sight in
the Galaxy. Recently, a very similar method has been applied to
determine the distances to some anomalous X-ray pulsars by Durant \&
van Kerkwijk (2006b).

Red clump giants have long been proposed as standard candles
(Paczynski \& Stanek 1998). These are core helium-burning giants with
a very narrow luminosity function that constitute a compact and
well-defined clump in an Hertzsprung-Russell diagram,
particularly in the infrared. Furthermore, as they are relatively
luminous, they can be identified to large distances from the Sun. The
location of stars of the same spectral type (i.e., of approximately
the same absolute magnitude) in a CMD
depends on distance and extinction. The effect of distance alone
shifts the stars to fainter magnitudes (vertically), while extinction
by itself makes the stars both redder and fainter (shifting them
diagonally in the CMD).

The absolute magnitude ($M_K$) and intrinsic color, $(J-K_{\rm s})_0$,
of the red clump giants are well established (Alves 2000; Grocholski
\& Sarajedini 2002; Salaris \& Girardi 2002; Pietrzy\'nski et
al. 2003). Here, we assume an absolute magnitude for the red clump
population of $-1.62\pm 0.03$ mag and an intrinsic color of $(J-K_{\rm
  s})_0$ = 0.7 mag. These values are consistent with the results
derived by Alves (2000) from the {\itshape Hipparcos} red clump data
and also with the results obtained by Grocholski \& Sarajedini (2002)
for open clusters. They are also in agreement with the intrinsic
colors for the red clump stars predicted by the Padova isochrones in
the Two Micron All Sky Survey (2MASS) system (Bonatto et al. 2004) for a
10 Gyr population of solar metallicity of $(J-K)_0$ = 0.68 mag.

The absolute magnitudes of the red clump stars have a small dependence
on metallicity (Salaris \& Girardi 2002). For our case, this effect is
negligible as there is no large metallicity gradient in the Galactic
disk (Ibata \& Gilmore 1995a, 1995b; Sarajedini, Lee, \& Lee
1995). However, the red clump star absolute magnitude in the $J$ band
is more sensitive to metallicity and age than the $K$ band, hence the
intrinsic color $(J-K)_0$ also depends on both the metallicity and age
(Salaris \& Girardi 2002; Grocholski \& Sarajedini 2002; Pietrzy\'nski
et al. 2003). For 4U\,1608$-$52 $ $(l=331.0$^\circ$, b=-0.9$^\circ$),
in particular, the maximum metallicity gradient expected in this field
produces a de-reddened color of $(J-K)_0=0.67\pm0.01$, considering the
recent results in the metallicity for red clump giants in the inner
Galaxy ([Fe/H]=-0.2 dex, Gonz\'alez-Fern\'andez et al. 2008), as well
as the predictions of the theoretical isochrones of Salaris \& Girardi
(2002) and Bonatto et al. (2004) for this range of metallicities.

In summary, the metallicity dependence leads to a systematic
uncertainty of 0.03 mag in the absolute magnitude of the red clump
stars and to 0.05 mag in the intrinsic color $(J-K)_0$ (a value that
also takes into account the dispersion around the mean color of the
red clump population). We will adopt these values when determining the
systematic uncertainties in the interstellar extinction along the line
of sight to 4U\,1608$-$52.

\subsection{Application to the Case of 4U~1608$-$52}
\label{apply}

To apply the red clump method in the Galactic field we are interested
in, we first build a ($J-K$, $K$) CMD. In the case of 4U 1608$-$52, we
use the data from 2MASS (Skrutskie et al. 2006) which is the most
complete database of NIR Galactic point sources available to date. The
2MASS JHK photometry is slightly shallower than that provided by the
more recent UKIDSS \emph{Galactic Plane Survey} (Lucas et
al. 2008). However, the latter survey has no coverage in the region
where 4U\,1608$-$52 lies, hence we rely on the 2MASS data only.

\begin{figure}[!h]
\centering

\includegraphics[width=10.0cm]{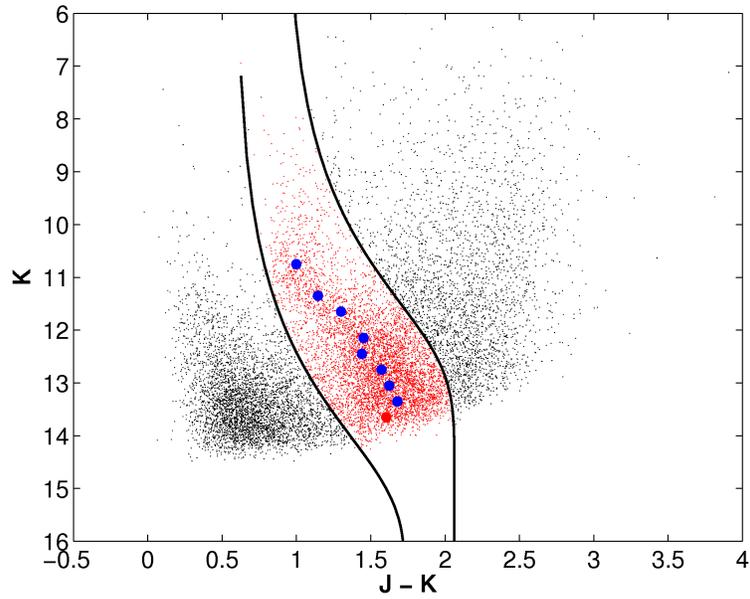}
\caption{NIR color-magnitude diagram for a 0.15 deg$^2$ field centered
  around 4U\,1608$-$52 ($l=330.9^\circ$ $b=-0.9^\circ$) and, taken
  from the 2MASS survey. The solid lines delimite the region where the
  red clump giants lie. Filled circles show points of maximum density
  of red clump stars for each individual magnitude bin.}
\label{Fig:CM1}
\end{figure}

In order to isolate the red clump sources, we use the theoretical
traces to define the limits of the K-giant branch on the CMDs (see
Figure \ref{Fig:CM1}). The traces are obtained for different stellar
types by using a double exponential approximation to the interstellar
extinction according to the updated ``SKY'' model (Wainscoat et
al. 1992). This very simple approach is used solely to facilitate the
separation of the K-giant branch and has no effect on the final
results. We then visually decide on the traces that are most useful in
isolating the K-giants. Our goal is to avoid any kind of contamination
in the red clump counts by other star populations, mainly dwarf stars
and M-giants (see Figure \ref{Fig:CM1}).

Once the optimal traces have been selected, we extract the giant stars
from the CMD and bin them according to their $K$ magnitudes. For each
magnitude bin, we construct count histograms by making horizontal cuts
(i.e., running color) through the CMDs at each range of $K$
magnitudes. We then fit a Gaussian function

\begin{equation}
f(x;A,\mu,\sigma)=A~\rm exp \left[ -\frac{(x-\mu)^{2}}{2\sigma^2} \right],
\label{ga}
\end{equation}

\noindent where $x$ is the binned $(J-K)$ color and $\mu$ is the color
of the peak, to the histogram to determine the position of the peak in
each cut.

\begin{figure}[!h]
\centering
\includegraphics[width=10.0cm]{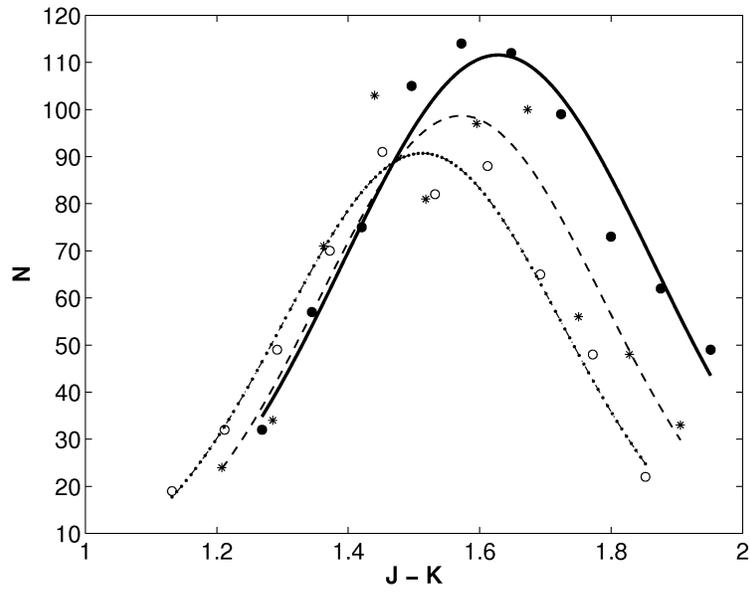}
\caption{Gaussian fits (lines) to the red clump counts (points) in
  three magnitude bins for the 0.15 deg$^{2}$ field around
  4U\,1608$-$52. The solid line (filled circles) corresponds to the
  12.6$<$m$_{K}$$<$12.9 bin, the dashed line (asterisks) to the
  12.3$<$m$_{K}$$<$12.6 bin, and the dot--dashed line (open circles)
  to the 12.0$<$m$_{K}$$<$12.3 bin.}
\label{Fig:gauss}
\end{figure}

Figure \ref{Fig:gauss} shows different Gaussian fits to the color
histograms obtained in three different magnitude bins for the field of
Figure \ref{Fig:CM1}. The identified maxima correspond to the red
clump stars since they are by far the most prominent population
(Hammersley et al. 2000; Cohen et al. 2000).

The fits provide simultaneously the magnitude m$_{K}$ and the color
$J-K$ for each maximum. We use these fits to determine the extinction
A$_{K}(m_{K})$ by tracing how the $J-K$ colors of the peaks of the red
clump counts change with $m_{K}$. We calculate the extinction for any
given $m_{K}$ using the measured $(J-K)$ of the peak, the intrinsic
mean color excess definition, and the interstellar extinction values
for $A_{J}$ and $A_{K}$, i.e.

\begin{equation}
A_{K}=c_e \times [(J-K)_{m_K}-(J-K)_{0}],
\label{ak}
\end{equation}

\noindent where $c_e$=0.657 (Rieke \& Lebofsky 1985; this value is 
consistent with the one found by Cardelli et al. 1989). We then obtain
the extinction $A_{K}(r)$ to a heliocentric distance $r$ for each
field by using the usual transformation for the distance along the
line of sight:

\begin{equation}
r=10^{[m_{K} - M_K + 5 - A_{K}(r)/5]}.
\label{r}
\end{equation}

We estimate the uncertainties in the extinction from the uncertainty
of the mean color of the Gaussian component

\begin{equation}
\sigma_{JKm} = \frac{\sigma}{1.52 \sqrt{N}},
\label{erJK}
\end{equation}

\noindent where $N=A\sigma \sqrt{2\pi}$ is the total number of red
clump stars in each magnitude bin. This uncertainty gives an indicator
of the statistical error only; to determine the total uncertainty in
the extinction, we also take into account the spread
$\sigma_{JK_{0}}$=0.05 mag in the intrinsic color, $(J-K)_0$, of the
red clump population due to metallicity effects and add both
uncertainties in quadrature:

\begin{equation}
\sigma_{A_K}^2 = \sigma_{JKm}^2 + \sigma_{JK0}^2
\label{sigak}
\end{equation}

We calculate the uncertainty in the distance along the line of sight
by means of Equation~(\ref{r}) from: $(i)$ the dispersion in the
absolute magnitude of the red clump population, $(ii)$ the
uncertainties in the extinction given by equation (\ref{sigak}), and
$(iii)$ the size of the binning in $m_K$ used to perform the Gaussian
fit of equation (\ref{ga}).

The only restriction that has to be taken into account when applying
this method is the contamination in the red clump counts due to dwarf
stars, which is important at $m_K>13$ for in-plane fields
(L\'opez-Corredoira et al. 2002). This imposes a limit in the distance
along the line of sight up to which we can extract the interstellar
extinction. For the field around 4U\,1608$-$52, this value is nearly
7.5 kpc, which turns out to be larger than the inferred distance to
the source.
 
\subsection{The Distance to 4U\,1608$-$52}

From the 2MASS archival data, we have extracted stars in the field of
view of 4U\,1608$-$52, using an interval of $\Delta l$=0.5$^\circ$ and
$\Delta b$=0.3$^\circ$, that gives a total number of 13202 stars. With
those stars, we measured the run of reddening with distance up to
$\sim$7.5 kpc.

In principle, we can increase the accuracy in the determination of the
extinction by enlarging the size of the region around the location of
4U\,1608$-$52. This would increase the number of red clump stars, thus
decreasing the uncertainty given by Equation (\ref{erJK}). At the same
time, it would allow a decrease in the bin size in magnitude $m_K$
that is used to perform the Gaussian fit of Equation (\ref{ga}),
yielding a better sampling of the extinction and a higher accuracy
when deriving the distance.

\begin{figure}[!h]
\centering
\includegraphics[scale=0.7]{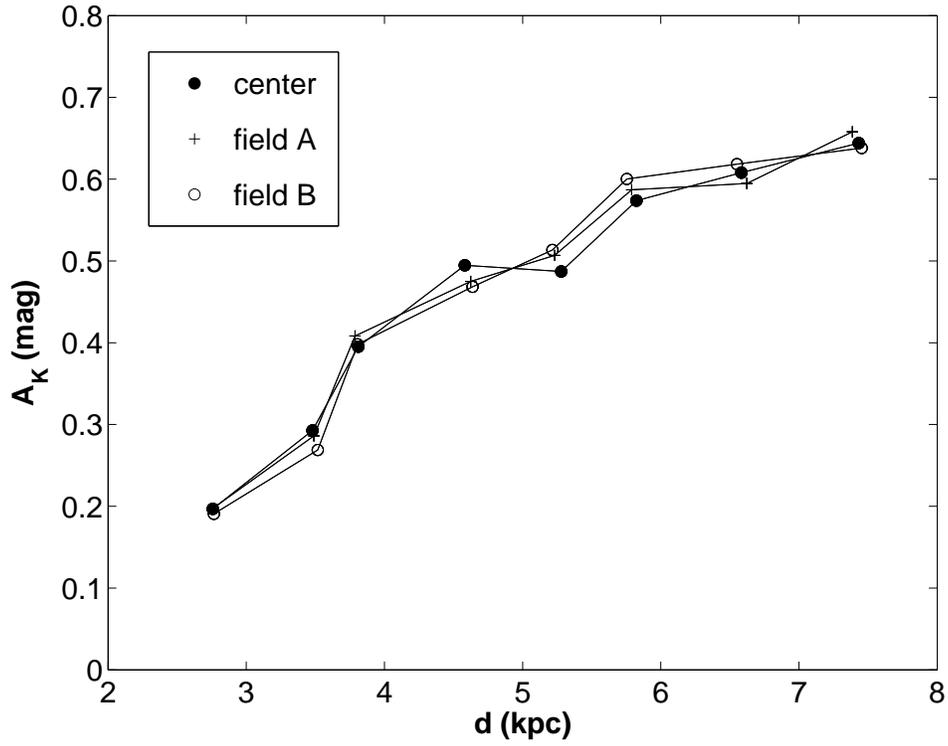}
\caption{K-band extinctions along the line of sight for the field 1.5
  deg$^{2}$ centered around 4U\,1608$-$52 (filled circles), and for
  two adjacent fields 0.5 deg away in longitude, labeled as 'field A'
  (pluses) and 'field B' (open circles). As it can be seen, there are
  no significant variations between three fields.}
\label{fields}
\end{figure}

To do this, however, we must first ensure that there are no
significant variations of extinction in the lines of sight around
4U\,1608$-$52. For this purpose, we collected 2MASS data in two
adjacent fields to 4U\,1608$-$52, each 0.5 degrees away in longitude
to the source (we will refer to them as `field A' and `field B'). We
find that there is no strong variation of extinction in those fields
with respect to the values obtained for the field centered on
4U\,1608$-$52 (see Figure \ref{fields}).

\begin{figure}[!h]
\centering
\includegraphics[width=12.0cm]{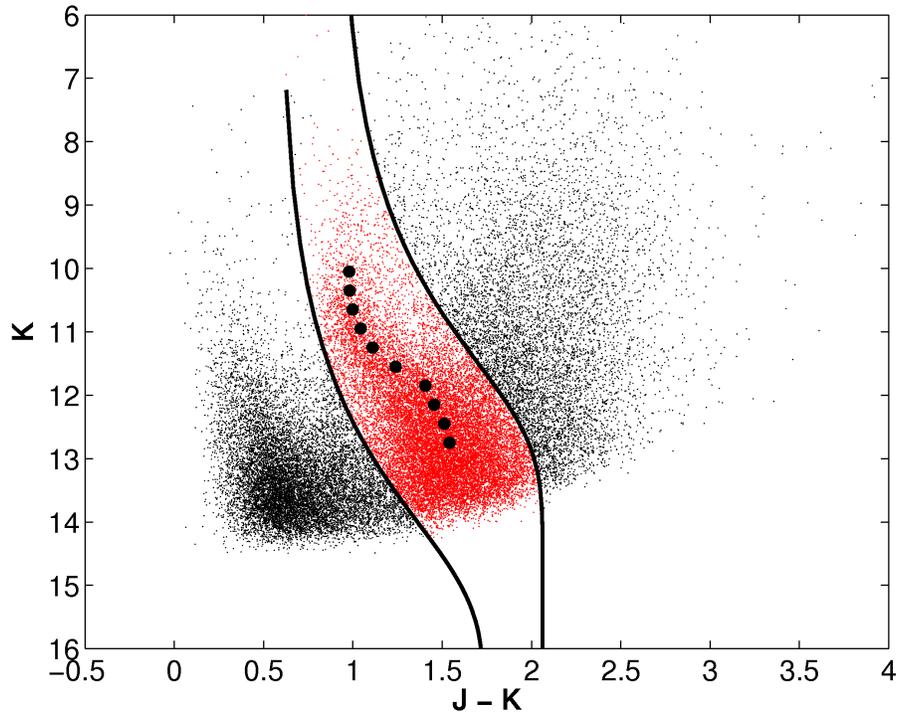}
\caption{Same as Figure \ref{Fig:CM1}, but now considering a
  0.45 deg$^2$ field centered around 4U\,1608$-$52 ($l=330.9^\circ$
  $b=-0.9^\circ$).}
\label{Fig:CM2}
\end{figure}

We merged all the data for the field centered on 4U\,1608$-$52, using
an interval for the extraction of 2MASS sources of $\Delta
l$=1.5$^\circ$ and $\Delta b$=0.3$^\circ$, hence covering an area of
$\sim$0.45 deg$^2$ in the sky. The resulting total number of stars
increased to 40548 and the uncertainties in the extinction estimates
decreased by a factor of 2.5. We show in Figure \ref{Fig:CM2} the
resulting CMD with the location for the maxima of the red clump
population derived following our method. In
Figure~\ref{distanceTOTAL}, we plot the inferred variation of
extinction along the line of sight to 4U\,1608$-$52.

To measure the distance to the binary, we compared the run of
extinction using the red clump stars to that obtained from the X-ray
spectrum of 4U\,1608$-$52. We denote the probability distribution over
extinction for the X-ray source as P$_{X}$(A$_{K}$) and the one
obtained for the red clump stars for each distance bin as
P$_{RC}$(A$_{K}|$D) and represent them by Gaussian functions. Because
there are no priors on either the extinction or the distance,
\begin{eqnarray}
P_{RC}(A_{K}|D) = P_{RC}(D|A_{K}).
\end{eqnarray}
We then calculated the total probability distribution by taking the
product of these two independent probability distributions and
marginalizing over the extinction:
\begin{eqnarray}
P(D)= \int P_{X}(A_{K}P_{RC}(D|A_{K})dA_{K}.
\end{eqnarray}

\begin{figure}[!h]
\centering
\includegraphics[width=10.0cm]{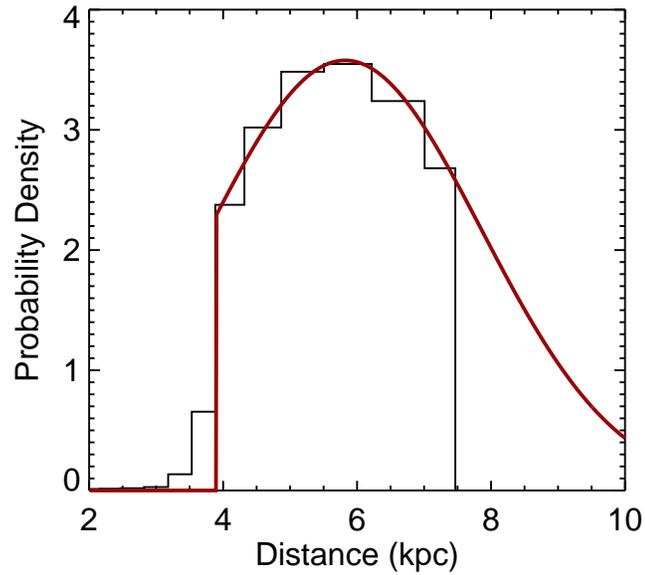}
\caption{The probability distribution over distance to the source
  4U~1608$-$52 and the best fit Gaussian model with a cut-off at 3.9
  kpc. The best fit distance is found to be 5.8$_{-1.9}^{+2.0}$ kpc.}
\label{dist_prob}
\end{figure}

In Figure~\ref{dist_prob}, we show the probability distribution over
distance calculated in this way. In order to find the distance with
the highest probability, we then fit the distribution with a Gaussian
function. Because there is a clear and sudden decrease in the
extinction below 3.9 kpc, we added a cutoff to the Gaussian function
below this distance. The best fit position of the Gaussian is
$5.8^{+2.0}_{-1.9}$~kpc, where the positive uncertainty reflects the
standard deviation of the Gaussian while the negative uncertainty
corresponds to the low-end cutoff.

\section{Modeling Type-I X-ray Bursts}

We analyzed RXTE proportional counter
array (PCA) archival observations of 4U\,1608$-$52, which totaled
1650~ks of observing time until 2007 June. We identified thermonuclear
(Type-I) X-ray bursts in the archival data using the X-ray burst
catalog of Galloway et al. (2008a). In total, we found 31 Type-I X-ray
bursts from this source.

Using the {\it seextrct} tool, we extracted 2.5$-$15.0~keV X-ray
spectra from all of the RXTE/PCA layers. Following the analysis in
\"Ozel et al. (2009), we used either the Science Event mode data
with the E\_125$\mu$s\_64M\_0\_1s configuration, which has a nominal
time resolution of 125 $\mu$s in 64 spectral channels, or we used {\it
Good Xenon} mode data which has 0.95 $\mu$s time resolution in 256
spectral channels, where available. We extracted spectra for 0.25,
0.5, 1, or 2~s time intervals, depending on the source count rate
during the burst. However, due to some data gaps during the
observations, in a few cases we had to use X-ray spectra integrated
over smaller exposure times. As in Galloway et al. (2008a) and \"Ozel
et al. (2009), we also used a 16~s spectrum prior to the burst as
background. We created separate response matrix files for each burst
using the PCARSP version 10.1.

We used XSPEC v12.4.0 (Arnaud 1996) for spectral analysis. We fit the
spectra with a blackbody function (bbodyrad model in XSPEC), fixing
the hydrogen column density to the value found from the analysis of
the high-resolution grating spectra (see Section~2).  We calculated
the bolometric X-ray fluxes using the Equation (3) of Galloway et
al. (2008a). All errors correspond to 1$-\sigma$ confidence levels of
our fits.

\begin{figure*}
\centering
\includegraphics[scale=0.50]{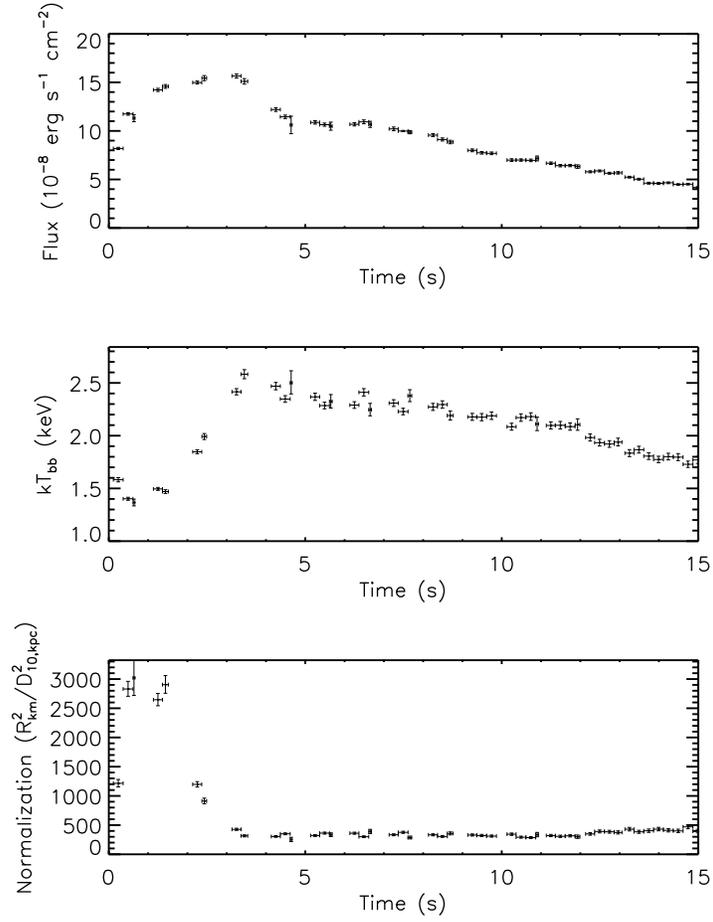}
\caption{Spectral evolution during the first 15 s of one
  the Eddington limited thermonuclear X-ray bursts (ID
  70059-03-01-000). The three panels show the flux, blackbody
  temperature and the normalization values together with their
  1$-\sigma$ statistical errors, respectively.}
\label{burst_curves}
\end{figure*}

\begin{figure*}
\centering
\includegraphics[scale=0.50]{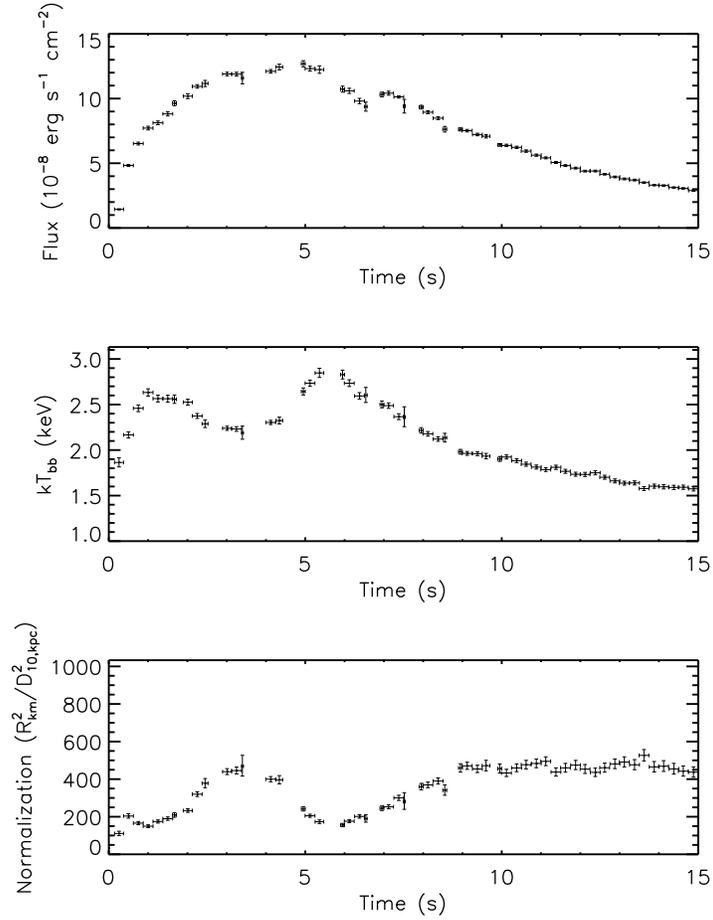}
\caption{Same as Figure \ref{burst_curves}, but for an X-ray burst (ID
  70059-01-26-00) that had previously been categorized as PRE
  (Galloway et al. 2008a) but did not meet our PRE identification
  criteria. Such bursts were not used in the analysis.}
\label{bad_burst_curves}
\end{figure*}

We identified the photospheric radius expansion bursts in this sample
by looking for {\em (i)} a clear difference in the apparent radius
between the first seconds of the burst and during the cooling tail and
{\em (ii)} a characteristic temperature evolution seen in PRE bursts.
The spectral evolution of an example of these bursts is given in
Figure~\ref{burst_curves}. We found a total of five such bursts, in
contrast to the 12 PRE bursts identified in Galloway et
al. (2008a). The difference arises from the fact that the latter study
categorized as PRE a number of bursts where the maximum radii reached
during bursts do not differ significantly from the radii during the
cooling tails.  We show in Figure~\ref{bad_burst_curves} an example of
such a case. We did not classify these as PRE bursts and, thus, did
not take them into account when determining the Eddington flux.

\begin{figure*}
\centering
\includegraphics[scale=0.70]{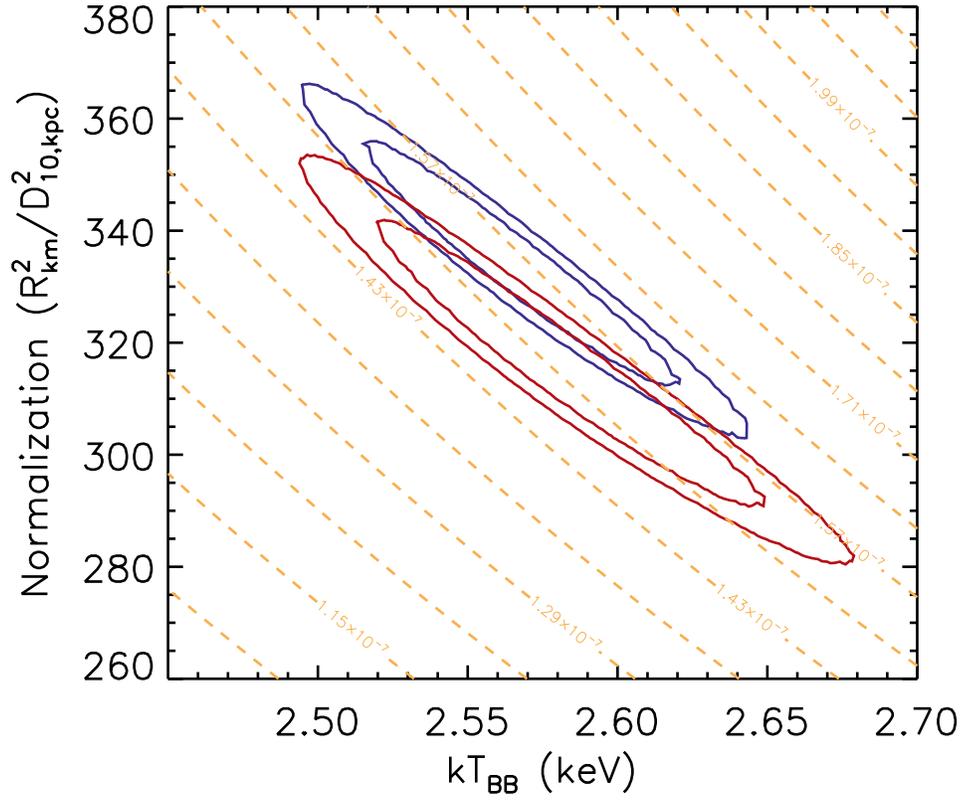}
\caption{The 1$-$ and 2$-\sigma$ confidence contours of the normalization
  and the blackbody temperature obtained from fitting the two PRE bursts
  during touchdown. The dashed lines show contours of constant bolometric
  flux.}
\label{touchdown}
\end{figure*}

We used the PRE burst sample shown in Table~\ref{tbursts} to measure
the touchdown fluxes.  We define the touchdown flux as the bolometric
flux measured when the normalization of the blackbody gets its lowest
and the temperature attains its highest value. In these bursts, the
peak flux during the burst and the touchdown flux are consistent with
each other to $\sim$5\% (cf., Galloway, \"Ozel, \& Psaltis
2008b). Table~\ref{tbursts} shows the measured touchdown
fluxes. Although two other bursts (with IDs 4 and 5) showed clear
evidence for a PRE event, data gaps during the touchdown instant
prevented us from determining and using their touchdown flux
values. We show in Figure~\ref{touchdown}) the confidence contours for
the color temperature and the normalization for the two PRE bursts
during touchdown. We found a combined best-fit value of (1.541
$\pm$0.065) $\times$10$^{-7}$ erg~cm$^{-2}$~s$^{-1}$ for 4U\,1608$-$52
for the touchdown flux.

The last spectroscopic quantity that we measured is the normalization
of the blackbody during the cooling tails of the bursts. This quantity
is related to the apparent blackbody radius and the distance through
$A = (R_{\rm{app}} / D_{\rm{10~kpc}})^{2}$, where $R$ is the apparent
radius in km and the $D_{\rm{10~kpc}}$ is the distance in units of
10~kpc. During the cooling tails of the Type-I X-ray bursts, the
blackbody radius attains a constant value, while the flux and the
temperature decrease. In order to find an average normalization for
each burst, we took the measurements in the 3$-$10~s time interval
after each burst started, when the apparent radius reaches a constant
value, while the signal-to-noise ratio remains high.  We present in
Table~\ref{nbursts} the blackbody normalizations for the four bursts
that showed this clear trend. Three out of the four measurements of
the blackbody normalization are within 1$-\sigma$, while the fourth
measurement is within 2$-\sigma$ of the rest. This implies that all
measurements are consistent with each other within their formal
uncertainties. Formally fitting these four values with a constant
results in a blackbody normalization of $324.6 \pm
2.4$~(km/10~kpc)$^2$.

\section{Measurement of the Neutron Star Mass and Radius}

The distance measurement to 4U\,1608$-$52, the touchdown flux, and the
blackbody normalization obtained by time-resolved X-ray spectroscopy
of the bursts provide the three observables that are needed to measure
independently the mass and the radius of the neutron star in this
binary. The observed spectral parameters depend on the stellar mass
and radius according to the equations (\"Ozel et al. 2009)

\begin{equation}
  F_{TD} = \frac{GMc}{\kappa_{es}D^2} \left(1-\frac{2GM}{Rc^{2}}\right)^{1/2}
\label{tdeq}
\end{equation}

\noindent and

\begin{equation}
  A = \frac{R^{2}}{D^2f_{c}^{4}} \left(1-\frac{2GM}{Rc^{2}}\right)^{-1},
\label{ampeq}
\end{equation}

\noindent where G is the gravitational constant, c is the speed of
light, $\kappa_{\rm es}$ is the opacity to electron scattering, and
$f_{c}$ is the color correction factor. We use the electron
scattering opacity $\kappa_{\rm es} = 0.20(1+X)$~cm$^{2}$~g$^{-1}$,
which depends on the hydrogen mass fraction $X$.

We assign independent probability distribution functions to the
distance $P(D)dD$, the touchdown flux $P(F_{TD})dF_{TD}$ and the
normalization $P(A)dA$ measurements, as well as to the possible range
of the hydrogen mass fraction $P(X)dX$ and the color correction factor
$P(f_c)df_c$. We then find the total probability density over the
neutron star mass $M$ and radius $R$ by integrating the equation

\begin{eqnarray}
&& P(D,X,f_c,M,R)dDdXdf_cdMdR=\frac{1}{2}P(D)P(X)P(f_c)P[F_{TD}(M,R,D,X)] \nonumber \\
&& \qquad \times P[A(M,R,D,f_c)]J(\frac{F_{TD},A}{M,R})dDdXdf_cdMdR\;,
\label{probcalc}
\end{eqnarray}

\noindent over the distance, the hydrogen mass fraction, and the color correction 
factor. Here, $J(F_{TD},A|M,R)$ is the Jacobian of the transformation
from the variables $(F_{TD}, A)$ to $(M,R)$.

In the absence of any observable constraints on the hydrogen mass
fraction, we assign a box-car probability distribution, allowing it to
cover the range from 0.0 to 0.7, i,e.,

\begin{eqnarray}
P(X)dX = \left \{
\begin{array}{lr}
\frac{1}{\Delta X} & {\rm if~} |X-X_{0}| \leq \Delta X/2 \\
0 & {\rm otherwise~}. \\
\end{array} \right.
\label{hfraction}
\end{eqnarray}

\noindent with $X_{0}$ = 0.35 and $\Delta X$=0.7. 

For the color correction factor $f_c$, we take a box-car probability
distribution covering the range $1.3-1.4$, so that

\begin{eqnarray}
P(f_c)df_c = \left \{
\begin{array}{lr}
\frac{1}{\Delta f_c} & {\rm if~} |f_c-f_{c0}| \leq \Delta f_c/2 \\
0 & {\rm otherwise~}, \\
\end{array} \right.
\label{fc}
\end{eqnarray}

where $f_{c0}  = 1.35$ and  $\Delta f_c =  0.1$. This is  motivated by
model atmosphere calculations shown  in Figure~\ref{fc} for a range of
neutron  star surface  gravity  strengths and  metallicities that  are
consistent  with the absence  of atomic  lines in  the high-resolution
X-ray spectra of bursters (see  Madej, Joss \& R\'o\'za\'nska 2004 and
Psaltis  \& \"Ozel  2009 for  the  details of  the calculations).  The
models of Majczyna et al. (2005) with solar and higher iron abundances
have  significant line  features that  have not  been detected  in any
high-resolution  burst  spectra;  nevertheless,  they result  in  very
similar  values of  the  color correction  factor.  Because the  color
correction is  applied to the temperature measured  during the cooling
tails of  bursts, where  the flux  has dropped to  less than  half the
Eddington flux  (for a given  surface gravity), the relevant  range of
values  is  indicated with  dashed  lines,  which  corresponds to  the
distribution quoted above.

All of the three measurements we use here are dominated by statistical
errors. For the distance, we use the parameters of the Gaussian
presented in Section 3, with a cutoff in the probability distribution
function below 3.9 kpc,

\begin{eqnarray}
P(D)dD = \left \{
\begin{array}{lr}
0 & {\rm for ~D < 3.9 ~kpc}  \\
\exp[-\frac{(D-D_{0})^{2}}{2\sigma^{2}_{D}}]  
& {\rm for ~D > 3.9 ~kpc.} 
\end{array} \right.
\label{dgauss}
\end{eqnarray}

\noindent appropriately normalized. 

Similarly, we assign a Gaussian probability distribution to the
blackbody normalization obtained from the bursts,

\begin{eqnarray}
P(A)dA =
\frac{1}{\sqrt[]{2\pi\sigma^{2}_{A}}}\exp[-\frac{(A-A_{0})^{2}}{2\sigma^{2}_{A}}]
\label{norgauss}
\end{eqnarray}

\noindent with A$_{0}$ = 324.6, $\sigma_{A}$ = 2.4 (km/10 kpc)$^{2}$.

For the touchdown flux, we take a Gaussian probability
distribution and use the best fit value and its standard deviation
given in Section 4.

\begin{eqnarray}
  P(F_{TD})dF_{TD} =
  \frac{1}{\sqrt[]{2\pi\sigma^{2}_{F}}}\exp[-\frac{(F_{TD}-F_{0})^{2}}{2\sigma^{2}_{F}}]
\label{ftdgauss}
\end{eqnarray}

The final distribution over the neutron star mass and radius is
obtained by inserting each probability distribution into Equation
(\ref{probcalc}) and integrating over the distance and the hydrogen
mass fraction. Figure~\ref{massradius} shows the 1 and 2 $-\sigma$
contours for the mass and the radius of the neutron star in
4U\,1608$-$52.

\section{Discussion}

\begin{figure}[!h]
\centering
\includegraphics[width=10.0cm]{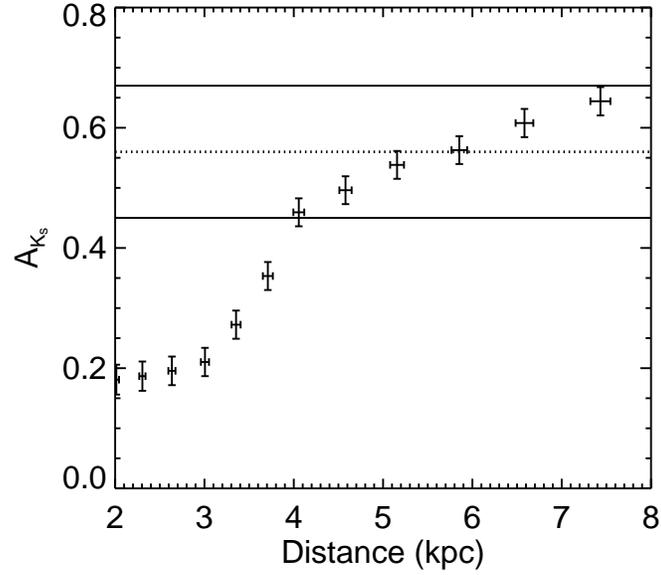}
\caption{The K-band extinction curve along the line of sight to
  4U~1608$-$52 using a 0.45 deg$^{2}$ field surrounding
  it. Overplotted are the A$_{K}$ value for 4U~1608$-$52 and its {\bf
  1$-\sigma$} statistical error.}
\label{distanceTOTAL}
\end{figure}

In this paper, we combined a distance measurement technique, applied
for the first time to a LMXB, with time-resolved spectroscopy of the
Type-I X-ray bursts in order to measure the mass and the radius of the
neutron star in the X-ray binary 4U\,1608$-$52. There are no previous
distance measurements to this binary. Our analysis led to a distance
of 5.8$^{+2.0}_{-1.9}$ kpc, which, together with the galactic coordinates of
the source, implies that 4U\,1608$-$52 is either on the far side of
the Scutum-Crux arm or in between the Scutum-Crux and Norma
arms. Between these two galactic arms, the increase in the extinction
is expected to be minor. Our analysis of the run of reddening with
extinction using the red clump giants along this line of sight showed
this expected trend. Because 4U\,1608$-$52 falls in a region between
the arms where the slope of the extinction curve shown in
Figure~\ref{distanceTOTAL} is very shallow, the errors in the distance
measurement, correspondingly, are relatively large.  Nevertheless, a
firm lower limit is achieved on the source distance.

\begin{figure*}
\centering
   \includegraphics[scale=0.80]{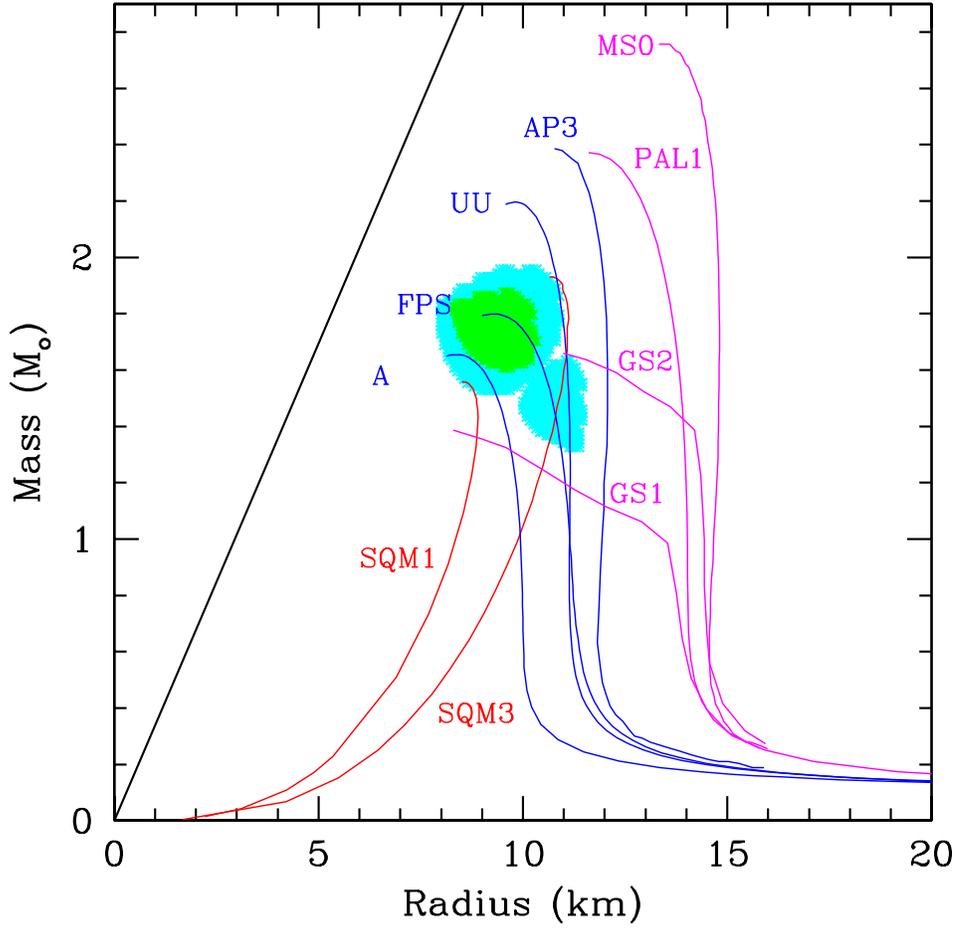}
   \caption{1$-$ and 2$-\sigma$ contours for the mass and the
     radius of the neutron star in 4U\,1608$-$52. The descriptions of
     the various equations of state and the corresponding labels can
     be found in Lattimer \& Prakash (2001).}
\label{massradius}
\end{figure*}

The 1- and 2-$\sigma$ confidence contours of mass and radius shown in
Figure~\ref{massradius} are based on the distance measurement and
spectral analysis of the X-ray bursts observed from this source. We
also find the individual probability density functions for mass and
radius by marginalizing $P(M,R)$ over radius and mass,
respectively. Fitting the resulting probability distributions with
Gaussian functions yields a mass of M~=~1.74~$\pm$0.14~M$_{\sun}$ and
and a radius of R~=~9.3$\pm$1.0~km, where the errors represent
1$-\sigma$ uncertainties.

There are numerous mass measurements and estimates of isolated and
binary neutron stars. Dynamical mass measurements allow, in some
cases, very precise determination of the neutron star masses (see
Thorsett \& Chakrabarty 1999 for a review). However, dynamical
measurements do not reveal any information about the neutron star
radii, which are better indicators of the neutron star equation of
state than are masses (\"Ozel \& Psaltis 2009).

Most attempts at measuring neutron star radii result in constraints on
the {\it combination} of neutron star radii and masses. Such
measurements have been carried out on globular cluster neutron stars
in binaries emitting thermally during quiescence, such as X7 in 47~Tuc
and others in $\omega$\ Cen, M~13, and NGC~2808 (Heinke et al.\ 2006;
Webb \& Barret 2007). These measurements have carved out large allowed
bands in the mass-radius plane, covering the mass-radius contours of
the present measurements.

Using multiple spectroscopic phenomena, on the other hand, leads to an
independent measurement of the stellar radius and mass. \"Ozel (2006)
used spectroscopic measurements of the Eddington limit and apparent
surface area during thermonuclear bursts, in conjunction with the
detection of a redshifted atomic line from the source EXO~0748$-$676,
to determine a mass of $M \ge 2.10 \pm 0.28~M_\odot$ and a radius $R
\ge 13.8 \pm 1.8~{\rm km}$. \"Ozel et al (2009) applied an approach 
similar to the one in the current paper to a neutron star binary
EXO~1745$-$248 in the globular cluster Terzan~5 and found tightly
constrained pairs of values for the mass and radius, which are
centered around $M=1.4~M_\odot$ and $R=11$~km or around
$M=1.7~M_\odot$ and $R=9$~km. The latter two radius measurements are
consistent with the one presented in the current paper to within
2$-\sigma$, and, therefore, several nucleonic equations of state are
consistent with both measurements.

The systematic and statistical errors on the measured masses and radii
of neutron stars do not yet allow the determination of a unique
equation of state for neutron star matter. However, by measuring a
sufficient number of neutron stars with different masses, it is
possible to trace out the mass and radius relation, which would then
allow a measurement of the pressure of cold ultradense matter at
densities beyond the nuclear saturation density (\"Ozel \& Psaltis
2009). In the present calculations, the largest uncertainty arises
from the distance measurement, due both to the particular location of
the binary in the Galactic disk as well as to the short exposure time
of the X-ray grating observation that yields a low signal-to-noise
spectrum. Future X-ray grating observations of will help decrease the
distance uncertainty and improve the mass and the radius measurement
as well.

\acknowledgements

We thank D. Psaltis and S. Bilir for very useful discussions and an
anonymous referee for numerous constructive suggestions that improved
the analysis. F.~\"O. acknowledges support from NSF grant AST
07-08640. This work made use of observations obtained with XMM-Newton,
an ESA science mission with instruments and contributions directly
funded by ESA Member States and NASA. This publication also made use
of data products from the 2MASS, which is a joint
project of the University of Massachusetts and the Infrared Processing
and Analysis Center/California Institute of Technology, funded by the
National Aeronautics and Space Administration and the National Science
Foundation.

\begin{deluxetable}{ccc}
  \tablecolumns{3} 
\tablewidth{210pt} 
\tablecaption{Log of all the XMM-Newton observations of 4U~1608$-$52.}  
\tablehead{Date & Obs. ID & Exp. Time \\ 
             & & (ks)} 
\startdata
  2002 Feb 13 & 0074140301& 2.9 \\
  2002 Feb 13 & 0074140101 & 17.2 \\
  2002 Feb 15 & 0074140201 & 16.9 \\
  2003 Mar 14 & 0149180201 & 7.2 \\
\enddata
\label{xmm_obs}
\end{deluxetable}

\begin{deluxetable}{ccccccc}
\tablewidth{495pt} 
\tablecolumns{7} 
\tablecaption{Best fit values of the parameters for the model}
\tablehead{Element & $\Gamma$ & Ampl. & Abs. Coef. & N & N$_{\rm{H}}$ & $\chi^2$/d.o.f.\\ 
& & $ (10^{-2})$ & $ $ &($10^{17}$ cm$^{-2}$) & ($10^{22}$ cm$^{-2}$) & }
\startdata
Mg & 3.80 $\pm$0.06 & 1.34 $\pm$0.02 & 0.053 $\pm$0.019 & 2.42 $\pm$0.87 &0.96 $\pm$0.34 & 1.1 (114 d.o.f.)\\ 
Ne & 7.69 $\pm$0.35 & 0.85 $\pm$0.02 & 0.34 $\pm$0.06& 9.65 $\pm$1.56 & 1.11 $\pm$0.18 & 1.4 (114 d.o.f.) \\ 
\enddata
\label{edgeres}
\end{deluxetable}

\begin{deluxetable}{cccc}
\tablewidth{310pt} 
\tablecolumns{4}
\tablecaption{Properties of the PRE bursts used to find the unabsorbed touchdown flux. Burst start
    times are adopted from Galloway et al. (2008a).}
  \tablehead{ID & Obs ID. & MJD & Touchdown Flux  \\
    & & & (10$^{-7}$ erg cm$^{-2}$ s$^{-1}$) } \startdata
  1& 30062-01-01-00  & 50899.58703 & 1.558$\pm$0.082 \\
  2& 70059-03-01-000 & 52529.17934 & 1.514$\pm$0.105\\
\enddata
\label{tbursts}
\end{deluxetable}

\begin{deluxetable}{cccc}
  \tablecolumns{4} \tablewidth{290pt} \tablecaption{Properties of the
    bursts used to find the normalization value during the cooling
    tail. Burst start times are adopted from Galloway et al. (2008a).}
  \tablehead{ID & Obs ID. & MJD & Normalization \\
    & & & (R$_{{\rm km}}$/D$_{\rm{10kpc}}$)$^{2}$ } \startdata
  2& 70059-03-01-000 & 52529.17934 & 326.7$\pm$4.7 \\
  3& 10093-01-03-000 & 50164.69334 & 330.2$\pm$4.9  \\
  4& 70059-01-20-00  & 52524.10157 & 325.8$\pm$5.4 \\
  5& 70059-01-21-00  & 52526.16006 & 317.0$\pm$4.5 \\
\enddata
\label{nbursts}
\end{deluxetable}

\begin{figure*}
\centering
   \includegraphics[scale=0.80]{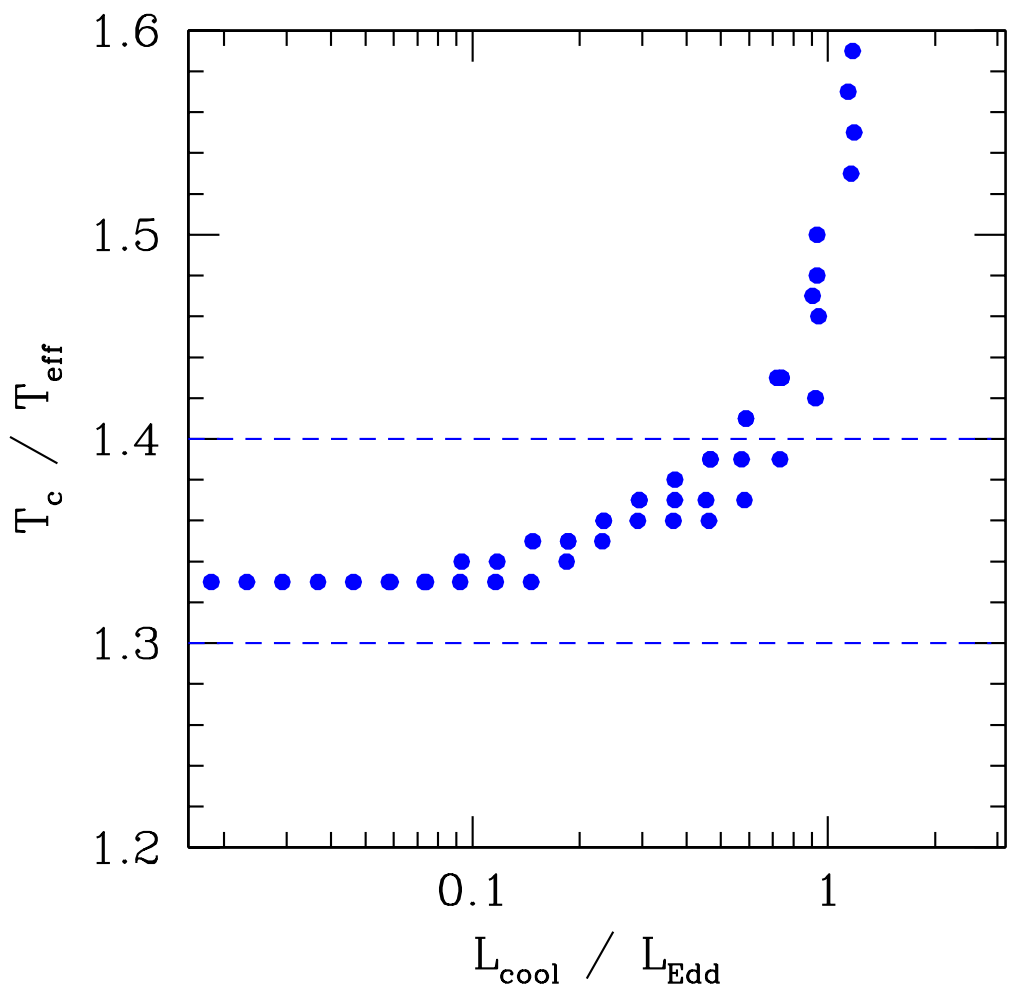} \caption{Color correction
   factor $f_c=T_c/T_{\rm eff}$ against the surface luminosity
   normalized to the Eddington luminosity. The values are obtained
   from model atmosphere calculations for a range of surface gravity
   strengths and for small metallicities that are consistent with the
   absence of atomic lines in the high-resolution X-ray spectra of
   bursters. The dashed lines indicate the range of color correction
   factors taken in the present work for the calculation of the
   apparent area of the neutron star during the cooling tails of
   bursts, where the luminosity drops to values less than $0.5 L_{\rm
   Edd}$.}
\label{fc}
\end{figure*}

\end{document}